\documentclass[prb,twocolumn,amsmath,amsfonts,showpacs]{revtex4-2}
\usepackage{graphicx}
\usepackage{epsf}
\usepackage{epstopdf}
\usepackage{graphicx}
\usepackage{tikz}
\usepackage{float}
\usepackage{amsmath,bm,upgreek}
\usepackage{nccmath}
\usepackage[mathscr]{euscript}
\usepackage{natbib}
\usepackage{hyperref}

\usepackage{amssymb}
\usepackage{graphicx}
\usepackage{epsf}
\usepackage{epstopdf}
\usepackage{graphicx}
\usepackage{tikz}
\usepackage{float}
\usepackage{amsmath}
\usepackage{nccmath}
\usepackage[mathscr]{euscript}

\begin{document}



\title{Quantum coherence in noise power spectrum in two quantum dots}
\author{Bogdan R. Bu{\l}ka}
\affiliation{Institute of Molecular Physics, Polish Academy of
Sciences, ul. M. Smoluchowskiego 17, 60-179 Pozna{\'n}, Poland}


\begin{abstract}
We present studies of quantum interference in a noise power spectrum in the system of two quantum dots (2QD) in a T-geometry. Performing the spectral decomposition we are able to separate local currents  and distinguish between the intra- and inter-level current correlation  contributions to the noise power spectrum. In particular, we analyzed the large bias regime and show that for a weak coupling of 2QD with the electrodes the noise power spectrum has dips at frequencies characteristic to  inter-level excitations.  For a strong coupling the electron transport changes its nature and the dynamics of the current correlations is different:  there are two coherently coupled relaxators with different relaxation frequencies. These two regimes of current dynamics are separated by a quantum critical point, in which the noise power spectrum shows a specific frequency dependence. In the linear response limit the noise power spectrum is related to the admittance, which shows characteristics different, due to quantum interference, for the weak and strong coupling case.

\end{abstract}



\maketitle

\section{Introduction}

In quantum optic experiments coherent properties of light are described by the coherent function of the first and the second order (or higher orders)~\cite{Glauber}. In analogy interference of electron waves were studied in transport through nanostructures, for example the Aharonov-Bohm effect~\cite{Webb}  and the Fano resonance~\cite{Gores}. These effects were observed in measurements of the differential conductance, which corresponds to the first order coherence function. The current correlation function (referred also as noise power) describes fluctuations of the currents and it corresponds with the second order coherence function. Problems of interference in the noise power spectrum have not  attracted much of interest \cite{Entin,Rothstein,Marcos}.

Here, we want to present studies of a frequency dependence of noise power, in particular, dynamics of coherent current correlations in the presence of quantum interference. Two quantum dots (2QD) in a T-shape geometry seems to be a suitable system for considerations, because the conductance shows the Fano resonance with a characteristic dip due to destructive interference of a travelling wave with an localized state. We
perform the spectral decomposition of noise power to get better insight into interference between local currents and to analyse dynamics of current correlations. The considerations are carried out for the large bias voltage as well as in the linear response limit, where the noise spectrum can be related to the admittance.

\section{Model description, current and noise calculations}

The considered 2QD system is described by the Hamiltonian
\begin{align}
\label{eq:ham}
H = \sum_{\sigma}\Big[t_{12} (c^{\dag}_{1\sigma}c_{2\sigma}+\text{h.c.})+ \sum_{i=1,2}\epsilon_i c^{\dag}_{i\sigma}c_{i\sigma}\Big]+\nonumber\\
\sum_{\substack{\alpha= L,R\\k,\sigma}} \epsilon_{\alpha k} c^{\dag}_{\alpha k\sigma} c_{\alpha k\sigma} +
\sum_{\substack{\alpha= L,R\\k,\sigma}}t_{\alpha} (c^{\dag}_{\alpha k\sigma}c_{1\sigma}+\text{h.c.})\;,
\end{align}
where the first term describes the 2QD system with a single level $\epsilon_{i}$ available at  the $i$-th QD and the interdot hopping $t_{12}$; the second term
describes electrons in the electrodes (in the left and the right one, $\alpha=L,R$), while the last term corresponds to the coupling of  the 1-st quantum dot to the electrodes, which is  described by the hopping parameter $t_{\alpha}$. In our model Coulomb interactions between electrons are neglected. To simplify notation the spin index $\sigma$ will be omitted. Notice, that the model (1) corresponds also to a single quantum dot with two levels,  with a weakly and a strongly coupled one, as well as to a dot coupled to fully polarized ferromagnets, where the dangling state would play the role of the minority spin.

 The current flowing from the $\alpha$-electrode to QD is expressed by
\begin{flalign}
I_{\alpha} =&
\frac{\imath 2e t_{\alpha}}{\hbar}\sum_{k}(\langle c^{\dag}_{\alpha k}d_{1}\rangle
-\langle d^{\dag}_{1}c_{\alpha k}\rangle)\nonumber\\ =&\frac{2e}{h}\int dE\;\mathcal{T}(E)[f_{L}(E) - f_{R}(E))]\;,
\end{flalign}
where we use the Green functions in the Keldysh form.  $f_{\alpha}$ denotes the Fermi distribution for electrons in the $\alpha$-electrode and the transmission is $\mathcal{T}(E)=4 \gamma_L \gamma_R/|z_1-t_{12}^2/z_2-\imath \gamma_N|^2$, $\gamma_N=\gamma_L+\gamma_R$, $z_i=E-\epsilon_i$. Here we used the wide-band approximation, for which the coupling of the 1-st dot with the $\alpha$-electrode is $\gamma_{\alpha}=\pi \rho_{\alpha} t^2_{\alpha}$, and $\rho_{\alpha}$ denotes the density of states.

To study the  noise power spectrum we use the Fourier transform of the correlation function~\cite{Aguado,Clerk}
\begin{eqnarray}\label{eq:sw-nons}
S_{\alpha \alpha'}(\omega)=\int dt\; e^{i\omega t} \langle\delta \hat{I}_{\alpha}(t)\delta \hat{I}_{\alpha' }(0)\rangle\;, \end{eqnarray}
for the currents in the $\alpha$ and  $\alpha'$ junction. The operator $\delta \hat{I}_{\alpha}(t)\equiv \hat{I}_{\alpha}(t) -I_\alpha $  describes current fluctuation from its average value.
The nonsymmetrized form of the noise spectrum is used in order to
 distinguish between emission ($\omega>0$) and absorption ($\omega<0$) of energy by a quantum detector (absorption and emission by the system)~\cite{Aguado,Clerk}.
Since we consider noninteracting electrons, the Wick's theorem can be applied to decouple two particle averages to products of single particle averages, which next are calculated by the Keldysh Green functions. The result is
\begin{align}
S_{\alpha \alpha'}(\omega)= \frac{4e^2}{h}\sum_{\eta, \eta'=L,R}\int dE\; \mathscr{S}_{\alpha\alpha'}^{\eta \eta'}(E,E+\hbar\omega)\nonumber\\
\times [1-f_{\eta}(E)]f_{\eta'}(E+\hbar\omega)\;, \end{align}
where the spectral density functions $\mathscr{S}_{\alpha\alpha'}^{\eta\eta'}(E,E+\hbar\omega)$ are expressed by the exact analytical formulae. We checked that the same results one can get by means of the scattering matrix method~\cite{Buttiker1992a,Buttiker1992b}. For example,
for the auto-correlation functions one gets their spectral density functions: $  \mathscr{S}_{LL}^{LL} =1-r_{LL}^* r_{LL}^{+}-r_{LL} r_{LL}^{+*} +\mathcal{R}\mathcal{R}^+$, $\mathscr{S}_{LL}^{RR}= \mathcal{T}\mathcal{T}^+$, $\mathscr{S}_{LL}^{RL}=\mathcal{T}(1-\mathcal{T}^+)$ and
$\mathscr{S}_{LL}^{LR}= \mathcal{T}^+(1-\mathcal{T})$. Here, we use the notation: $\mathcal{T}\equiv t_{LR}t_{LR}^*$, $t_{LR}=\imath 2\sqrt{\gamma_L \gamma_R}/(z_1-t^2_{12}/z_2 + \imath \gamma_N)$; $\mathcal{R}\equiv 1-\mathcal{T}= r_{LL}r_{LL}^*$, $r_{LL}=[z_1-t^2_{12}/z_2-\imath (\gamma_L- \gamma_R)]/(z_1-t^2_{12}/z_2 + \imath \gamma_N)$, and $\mathcal{T}^+=\mathcal{T}(E+\hbar\omega)$, $\mathcal{R}^+=\mathcal{R}(E+\hbar\omega)$, $r_{LL}^+=r_{LL}(E+\hbar\omega)$.

\section{Analysis of the noise power spectrum}

\subsection{Large bias voltage}

In the large bias voltage regime both the states, the bonding and antibonding one, are in the transport window. Moreover, we assume moderate frequencies, $\hbar|\omega|< e|V|$, thus the Fermi distribution functions $f_{L}(E)=f_L(E+\hbar\omega)=1$,  $f_{R}(E)=f_R(E+\hbar\omega)=0$, and the currents are unidirectional, there are not backscattering processes.  In this case the current correlation function is expressed only by the shot noise term, that one related to the spectral density function $\mathscr{S}_{LL}^{RL}(E,E+\hbar\omega)$. Fig.1 presents the plots of $S_{LL}(\omega)$ for different couplings to the electrodes. The central dip is related with anti-bunching of scattered electrons and its minimal value $S_{LL}(0)/2eI=1/2$ is characteristic for a resonant transport~\cite{BlanterButtiker}.
There are two extraordinary side-dips which origin will be investigated below in details.

\begin{figure}
\includegraphics[width=1.\linewidth,clip]{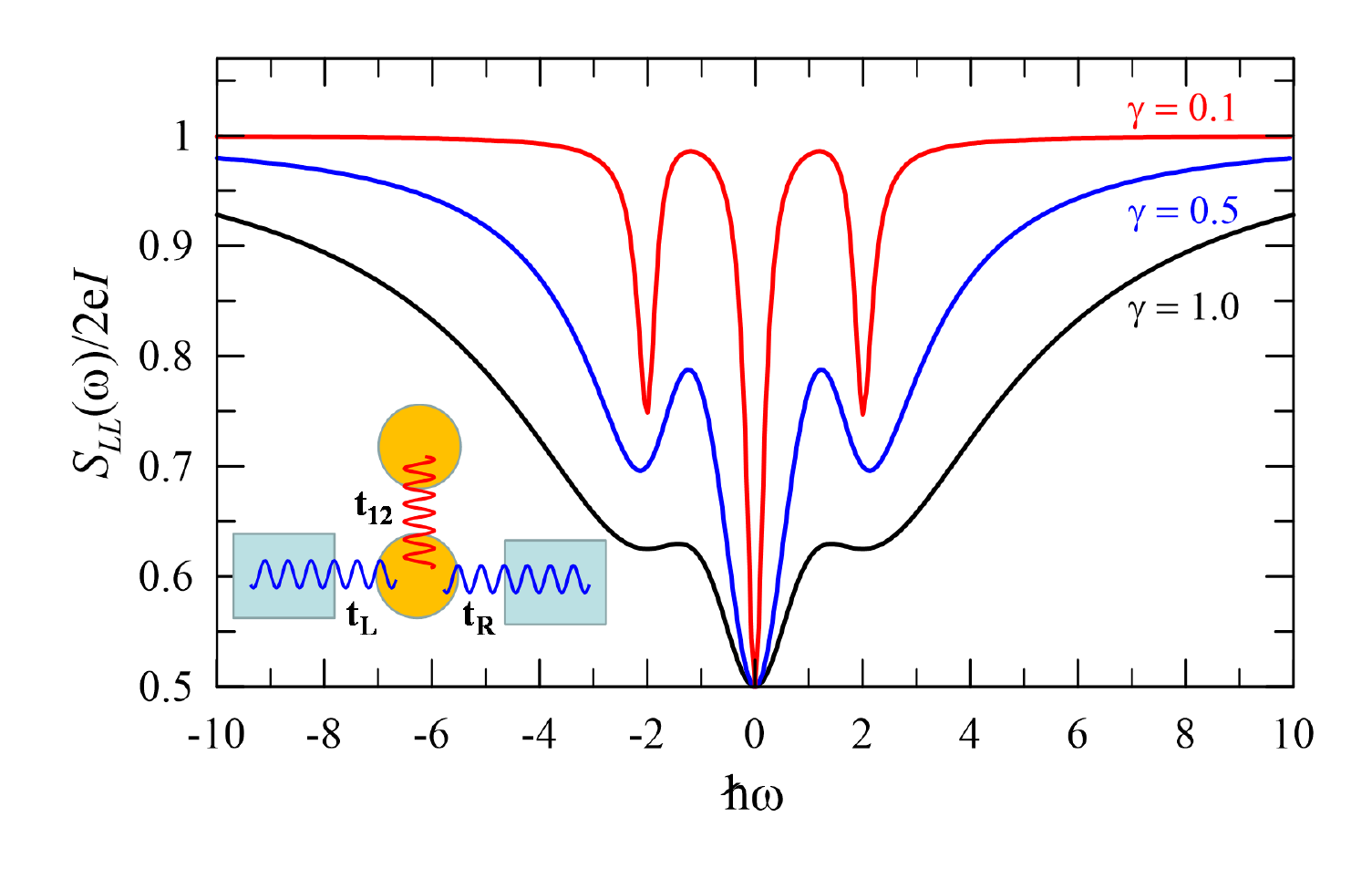}
\caption{Frequency dependence of the current correlation $S_{LL} (\omega)$  in the large voltage bias regime for different couplings, $\gamma_L=\gamma_R=\gamma=0.1$, $0.5$  and $1.0$ (the red, blue and black curve, respectively). The dot levels are taken: $\epsilon_1=\epsilon_2=0$ and the inter-dot hopping parameter $t_{12}=-1$, which is taken as unity. The plots are normalized to $2eI$, with the current $I=(2e/\hbar)2\gamma_L\gamma_R/\gamma_N$. The inset shows the considered 2QD system.}\label{fig1}
\end{figure}

To analyse the spectrum of noise power we perform the spectral decomposition. The current $I=I^-+I^+$ is decomposed into the currents flowing through the
bonding and anti-bonding level, $\epsilon_{\mp}=\epsilon\mp\Delta$, respectively. This decomposition enables us for insight into interference of electron waves travelling through different optical paths as well as into correlations of local currents~\cite{Bulka2019}.  Here, to simplify presentation we consider the symmetric system with $\gamma_L=\gamma_R=\gamma$ and $\epsilon_1=\epsilon_2=\epsilon$, for which the interlevel separation is
$2\Delta=2\sqrt{t^2_{12}-\gamma^2}$. The dimensionless conductances can be written as: $\mathcal{G}^{\mp}=\mp (E-\epsilon)\gamma^2/\{\Delta[(E-\epsilon_{\mp})^2+\gamma^2]\}$.
In similar way we can decompose the noise power spectrum: first the spectral
density function $\mathscr{S}_{LL}^{RL}(E,E+\hbar\omega)$ is decomposed, and next one integrates using the residue theorem. The result is
\begin{align}
&S_{LL}(\omega)=2e I+ \sum_{\nu,\nu'=\pm}S_{LL}^{\nu\nu'}(\omega)\nonumber\\
&\qquad\quad\; = \frac{4e^2}{\hbar}  \gamma\bigg\{1 - \frac{2\gamma^2(\Delta^2+\gamma^2)}{ \Delta^2 (\hbar^2 \omega^2+4 \gamma^2)} + \nonumber\\
& \frac{\gamma^2[\gamma^2 +\Delta( \Delta-\hbar\omega)]}{ \Delta^2 [(2\Delta-\hbar\omega)^2+ 4\gamma^2]}
+ \frac{\gamma^2[\gamma^2 +\Delta( \Delta+\hbar\omega)]}{ \Delta^2 [(2\Delta+\hbar\omega)^2 +4\gamma^2]}\bigg\},
\end{align}
where the first term corresponds to the Schottky noise, which is frequency independent and describes uncorrelated transfers of particles. The other terms are frequency dependent; the second one (in the curly bracket) describes the intra-level current correlations, $S_{LL}^{\mp\mp}$, while the third and the fourth one describes the inter-level current correlations, $S_{LL}^{\mp\pm}$, with emission and absorption of the energy $\hbar\omega=2\Delta$, respectively. Let us notice that the auto- and the cross-correlation functions are related $S_{LL}(\omega)=2e I+S_{LR}(\omega)$ in the considered large voltage limit.

The above analysis has been performed for the weak coupling with the electrodes, however, for the strong coupling $\delta^2=\gamma^2-t^2_{12}>0$ one gets different results. The spectral decomposition of the conductance gives: $\mathcal{G}^{\mp}=\mp \gamma(\gamma\mp\delta)^2/\{\delta[(E-\epsilon)^2+(\gamma\mp\delta)^2]\}$, which means that there are two transport channels through a single level $\epsilon$. The noise power spectrum is expressed as
\begin{align}
&S_{LL}(\omega)= \frac{4e^2}{\hbar}  \gamma\bigg\{1 +\frac{2\gamma^2(\gamma^2-\delta^2)}{ \delta^2 ( \hbar^2\omega^2 +4 \gamma^2)} \nonumber\\
&- \frac{\gamma(\gamma-\delta)^3}{ \delta^2 [\hbar^2\omega^2+4(\gamma-\delta)^2]} - \frac{\gamma(\gamma+\delta)^3}{\delta^2 [\hbar^2\omega^2+4(\gamma+\delta)^2]}\bigg\}\;,
 \end{align}
which is composition of dynamics of two relaxators. The third and the fourth term describes auto-correlations of the relaxators with the characteristic relaxation frequency: $\omega^-=(\gamma-\delta)/\hbar$ and $\omega^+=(\gamma+\delta)/\hbar$, respectively. The second term corresponds to inter-relaxator correlations.

These results show that the coherent dynamics and relaxation processes for the weak and the strong coupling to the electrodes are different.  Two regimes are separated by the quantum critical point, at $\gamma=|t_{12}|$, in which the noise power spectrum is
\begin{align}
&S_{LL}(\omega)= \frac{4e^2}{\hbar}\gamma \bigg\{1-\frac{4\gamma^2}{\hbar^2\omega^2+4\gamma^2}\nonumber\\
&- \frac{4\gamma^2(\hbar^2\omega^2-4\gamma^2)}{(\hbar^2\omega^2+4\gamma^2)^2} -\frac{8\gamma^4(-3\hbar^2\omega^2+4\gamma^2)}
{(\hbar^2\omega^2+4\gamma^2)^3}\bigg\}.
\end{align}

\subsection{Equilibrium noise and fluctuation-dissipation theorem}

\begin{figure}
\includegraphics[width=1.\linewidth,clip]{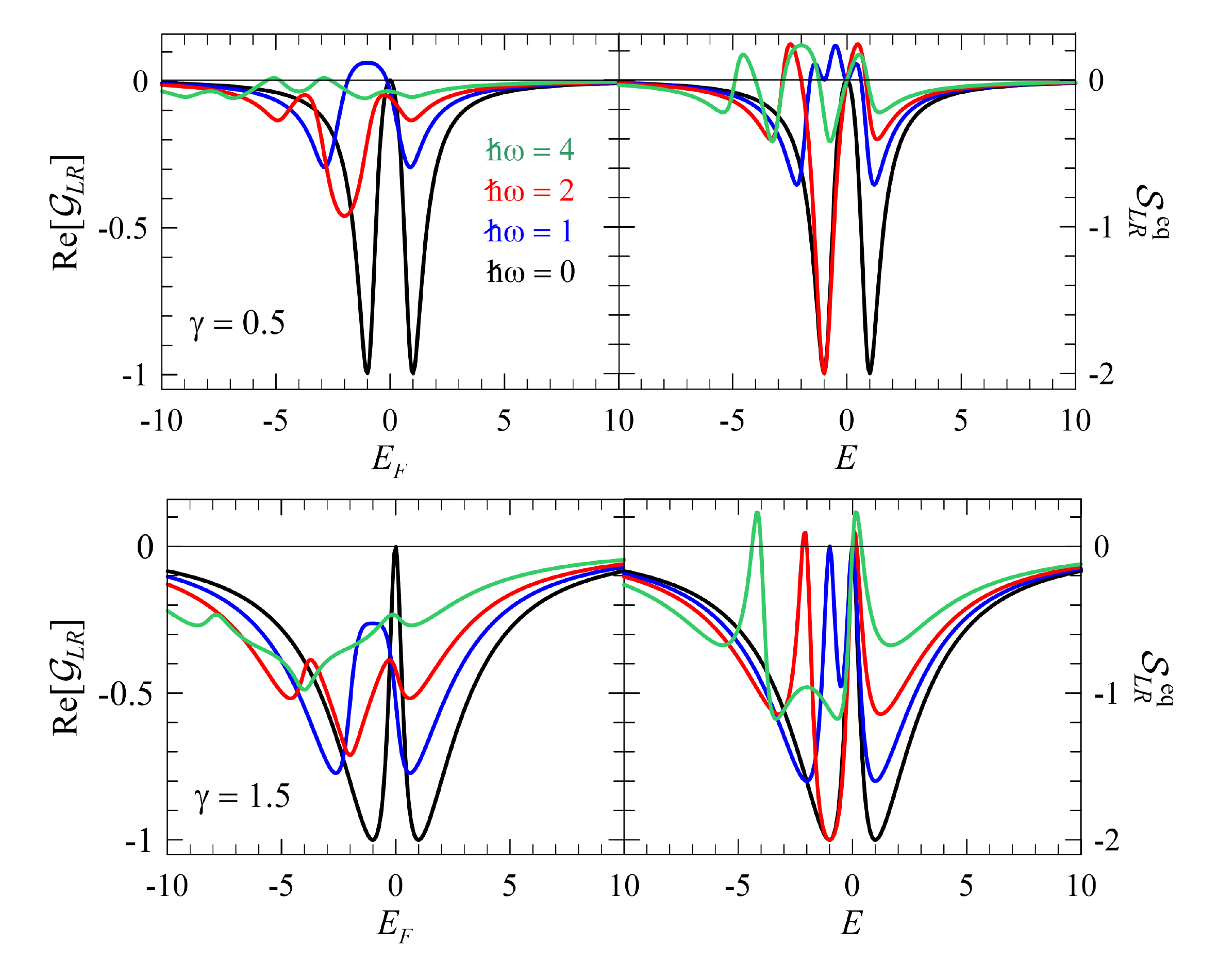}
\caption{Real part of the admittance $\mathcal{G}_{LR}$  (in units of $2e^2/h$) and the density of the current cross-correlation function $\mathscr{S}_{LR}^{\text{eq}}$ plotted vs the Fermi energy $E_F$ and the single electron energy $E$ for different $\hbar\omega=$ 0, 1, 2, 4, as well as for the weak and strong coupling to the electrodes, $\gamma=0.5$ and $\gamma=1.5$, respectively. Temperature is taken $T=0$ and the other parameters are the same as in Fig.1.}\label{fig2}
\end{figure}

For a small bias voltage one can use the the Kubo theory of linear response  and establish the relation of the noise power $S_{\alpha\alpha'}$ with  the admittance
$\mathcal{G}_{\alpha\alpha'}$ by the fluctuation-dissipation theorem
\cite{Nazarov,Fu,Buttiker1993}
\begin{align}\label{admit}
S_{\alpha\alpha'}(\omega)-S_{\alpha'\alpha}(-\omega)
= 4\hbar\omega\Re[\mathcal{G}_{\alpha\alpha'}(\omega)]=
\nonumber\\
\frac{4e^2}{h}\int dE\; \mathscr{S}_{\alpha\alpha'}^{\text{eq}}(E,E+\hbar\omega) [f(E+\hbar\omega)-f(E)]\;,
\end{align}
where the real part of $\mathcal{G}_{\alpha\alpha'}$ is responsible for dissipation. Here, we denoted
$\mathscr{S}_{\alpha\alpha'}^{\text{eq}}\equiv\sum_{\eta,\eta'=L,R} \mathscr{S}_{\alpha\alpha'}^{\eta\eta'}$, which can be explicitly expressed as  $\mathscr{S}_{LL}^{\text{eq}}= 2-r_{LL}^* r_{LL}^{+}-r_{LL} r_{LL}^{+*}$ and $\mathscr{S}_{LR}^{\text{eq}}= -t_{LR}^* t_{LR}^{+}-t_{LR} t_{LR}^{+*}$ for the auto- and cross-correlations, respectively.
The admittance can be also calculated directly using either the scattering matrix or the Green function method \cite{Anantram, Shevtsov}.

Fig.2 presents the plots for the current crossed correlation: the admittance  $\Re[\mathcal{G}_{LR}]$ and the density function $\mathscr{S}_{LR}^{\text{eq}}$ calculated at zero temperature $T=0$ and for
different frequencies. For $\hbar\omega=0$, one can see that $-\Re[\mathcal{G}_{LR}]$ is equal to the conductance $\mathcal{G}$ with its
characteristic shape due to destructive interference. The conductance describes a first-order interference process as in a Young's experiment. However, for finite frequencies the admittance contains also second-order
interference processes (two-particle correlations)~\cite{Glauber} as in a Hanbury Brown-Twiss intensity interferometry.
At $T=0$ the noise power spectrum is nonzero for positive $\hbar\omega>0$,
i.e. the 2QD system absorbs energy. All electron waves, with energy $E\in [E_F,E_F+\hbar\omega]$, contribute to the admittance, which shows different
dependence for the weak and strong coupling, $\gamma=0.5$ and $\gamma=1.5$ (the upper and lower panel in Fig.2). Notice that $\Re[\mathcal{G}_{LR}]$ can
be positive in some energy range for the weak coupling case, what means that electrons can be bunched.
Interference is better seen in the density function $\mathscr{S}_{LR}^{\text{eq}}$. In particular, its peaks can be associated  with interference processes, which are different for both coupling cases.
   One can perform also the spectral decomposition of the frequency dependent admittance (similarly as in the previous section for the shot noise).
The result is more complex; one finds intra- and inter-channel intensity correlations, different for the weak and strong coupling, but there are also terms describing the second-order interference processes which are associated with the positive values of $\mathscr{S}_{LR}^{\text{eq}}$, i.e. they are responsible for electron bunching.

\section{Summary and final remarks}

We derived, using the Keldysh Green functions, exact formulae for the current and  the frequency dependent current-current correlation functions, Eq.(5)-(7), in the 2DQ system, where the Fano resonance occurs.
For the high voltage bias the shot noise dominates, which shows the particle nature of the electron transport. Performing the spectral decomposition we were able to separate the currents flowing through the bonding and the antibonding state, and distinguish between the intra- and inter-level current correlation contributions to the noise power spectrum. In particular, we showed that for a weak coupling with the electrodes the noise spectrum has dips at frequencies characteristic to inter-level excitations and the corresponding current correlations are negative.  When the coupling with the electrode is larger than the separation between the states, the electron transport changes its nature.
The dynamics of the current correlations is different:  there are two coherently coupled relaxators with different relaxation frequencies.
These two regimes of current dynamics are separated by a quantum critical point.

For a small voltage bias we used the linear response theory in which the noise power spectrum is related to the admittance. Although for zero-frequency one recovers the result for the conductance with a dip due to destructive interference, but for finite-frequencies the admittance is more complex.
There are the first-order as well as the second-order interference processes
for many electron waves (from the energy window $[E_F,E_F+\hbar\omega]$). Spectral decomposition shows that intensity correlations are responsible for electron bunching which is well seen in $\Re[\mathcal{G}_{LR}]$ in the weak coupling case (Fig.2).

The interplay of quantum coherence and Coulomb correlations in the noise power spectrum remains an open objective, an interesting one but difficult for calculations because one should treat properly many body correlations.

Our theoretical predictions can be verified by measurement of noise power spectrum in an active quantum detector  coupled via an on-chip resonant circuit to the quantum dot system \cite{Clerk} or with a multilevel quantum sensor \cite{Sung2021} for more complex systems. An open issue is a sensitivity of these devices to detect dynamics of the current correlations with the inter-channel absorption/emission of energy.

Let us also mention on a single electron source technique~\cite{Glattli} in which single particle wavepackages (levitons) propagate ballistically along edge quantum Hall channels. Within this technique levitons can be guided in an optics like setup, both in time or energy, which enables to perform tests of electron interferometry and entanglement. A key issue to verify our predictions will be fabrication a setup with a chiral edge channel coupled to a localized state in which the Fano resonance will appear.

\acknowledgments{
The research was financed by National Science Centre,
Poland — project number 2016/21/B/ST3/02160.}

\bibliographystyle{apsrev4-2}
\bibliography{2qd-splitter}
\end{document}